\documentclass [prl,twocolumn,superscriptaddress,showpacs,preprintnumbers] {revtex4}
\usepackage[]{amssymb}

	\usepackage{amsmath}%,amssymb} 
	\usepackage{makeidx}
	\usepackage{amsfonts}
	\usepackage[ansinew]{inputenc}
	\usepackage[usenames,dvipsnames]{pstricks}
	\usepackage{subfigure}
	\usepackage{epsfig}
	\usepackage{graphicx}% Include figure files
	\usepackage{dcolumn}% Align table columns on decimal point
	\usepackage{bm}
	\usepackage[colorlinks,hyperindex]{hyperref}
	\usepackage{epstopdf}	
	\usepackage{float}
	\usepackage{perpage}

	\newsavebox\CBox
	\newcommand\hcancel[2][0.5pt]{%
	 \ifmmode\sbox\CBox{$#2$}\else\sbox\CBox{#2}\fi%
	 \makebox[0pt][l]{\usebox\CBox}%  
	 \rule[0.5\ht\CBox-#1/2]{\wd\CBox}{#1}}

	\MakeSorted{figure}
	\MakeSorted{table}
	\hypersetup
	{
		colorlinks,%
		citecolor=black,%
		linkcolor=black,%
		urlcolor=black,%
	}

	\newcommand{\vet}[1]{\mathbf #1 }

	\newcommand{\um}[1]{\mathrm{#1}}
	\newcommand{\nd}{\noindent}

\makeindex

\begin{document}

\title{Temperonic Crystal: a superlattice for temperature waves in graphene}

\author{Marco Gandolfi}
\email{marco.gandolfi@ino.cnr.it}
\affiliation{CNR-INO (National Institute of Optics), Via Branze 45, 25123 Brescia, Italy}
\affiliation{Department of Information Engineering, University of Brescia, Via Branze 38, 25023 Brescia, Italy}

\author {Claudio Giannetti}
%\email{claudio.giannetti@unicatt.it}
\affiliation{Department of Physics, Universit{\`a} Cattolica del Sacro Cuore, Via Musei 41, 25121 Brescia, Italy}
\affiliation{Interdisciplinary Laboratories for Advanced Materials Physics (I-LAMP), Universit{\`a} Cattolica del Sacro Cuore, Via Musei 41, 25121 Brescia, Italy}

\author {Francesco Banfi}
\email{francesco.banfi@univ-lyon1.fr}
\affiliation{FemtoNanoOptics group, Universit\'{e} de Lyon, CNRS, Universit\'{e} Claude Bernard Lyon 1, Institut Lumi\`{e}re Mati\`{e}re, F-69622 Villeurbanne, France}

\begin{abstract}
The temperonic crystal, a periodic structure with a unit cell made of two slabs sustaining temperature wave-like oscillations on short time-scales, is introduced. The complex-valued dispersion relation for the temperature scalar field is investigated for the case of a localised temperature pulse. The dispersion discloses frequency gaps, tunable upon varying the slabs thermal properties. Results are shown for the paradigmatic case of a graphene-based temperonic crystal. The temperonic crystal extends the concept of superlattices to the realm of temperature waves, allowing for coherent control of ultrafast temperature pulses in the hydrodynamic regime at above liquid nitrogen temperatures.
\end{abstract}

\maketitle

%%%%%%%%%%%%%%%%%%%%%%%%%%

Coherent control of wave-like phenomena via metamaterials is driving, ever since four decades, a technological revolution in fields ranging from electronics, photonics, to phononics \cite{istrate2006photonic,weisbuch2014quantum,maldovan2013sound,wegener2013meta,zheludev2012metamaterials,matsuda2015fundamentals}. 
Although temperature has been historically considered as the paradigmatic example of an incoherent  field, undergoing diffusive as opposed to wave-like propagation, on short space and time scales Fourier law fails \cite{Minnich2011,Johnson2013,Hoogeboom2015,chen2018,Frazer2019} and the possibility for temperature wave-like propagation sets in \cite{Zhang2020,hua2020Space,beardo2020,huberman2019observation,beardo2020phonon}.
The ultimate goal is to devise metamaterials, addressed as temperonic metamaterials, enabling coherent control of temperature oscillations arising in the hydrodynamic heat transport regime and operating at above liquid nitrogen temperature.\\
\indent To this end the Temperonic Crystal (TC), a periodic structure based on a unit cell (u.c.) composed of two slabs sustaining temperature wave-like oscillations on short time-scales, is here introduced as an archetypal example. The TC is the analogue, for the temperature case, of electronic, phononic and photonic superlattices. The complex-valued dispersion relation for the temperature scalar field in TCs is investigated for the case of a localised temperature pulse. The dispersion discloses frequency gaps, tunable upon varying the slabs thermal properties and dimensions, serving, for instance, as a frequency filter for a temperature pulse triggered by an ultra-short laser. Results are shown for the realistic case of a graphene-based TC, graphene having recently shown to sustain temperature wave-like oscillations at above liquid nitrogen temperatures \cite{huberman2019observation,gandolfi2017emergent,Lee2015,ding2018,cepellotti2015phonon}.
The TC serves as a conceptual building block to manipulate nanoscale heat transfer by exploiting interference effects.\\
\indent In order to account for non-Fourier heat transport, occurring on short time and length scales, the present work relies on the Dual-Phase-Lag (DPL) model \cite{tzou1995unified, tzou1995experimental,tzou2014macro}. The DPL introduces a causality relation between the onset of the heat flux, $\vet{q}$, and the temperature gradient, $\nabla T$: $\vet{q}\left(\vet{x},t+\tau_q\right)=-\kappa_T\ \nabla T \left(\vet{x},t+\tau_T\right)$,
\nd where $\kappa_T$ is the thermal conductivity and $\tau_T$ and $\tau_q$ are positive build-up times for the inception of $\nabla T$ and $\vet{q}$ respectively \cite{Ordonez-Miranda2010_MRC}, which microscopic attribution varies according to the specific physical system \cite{gandolfi2019accessing}. In the following, 1D propagation will be considered. Expanding the DPL model to first order \cite{sobolev1997local,Al-Nimr2000}, while imposing local conservation of energy at time $t$, yields:  
\begin{equation}
\left\{\begin{array}{l}
\displaystyle{q+\tau_q\frac{\partial q}{\partial t}\simeq-\kappa_T\ \frac{\partial T}{\partial x}-\kappa_T \tau_T\ \frac{\partial^2 T}{\partial x\partial t}}\\
\displaystyle{ C\ \frac{\partial T}{\partial t}=-\frac{\partial q}{\partial x}}\\
\end{array}
\right.
\label{cons_en_DPLM}
\end{equation}

which may be recast as:
\begin{equation}
\left( {\tau_q\over \alpha}\right)\frac{\partial^2 T}{\partial t^2}-\frac{\partial^2 T}{\partial x^2}+{1\over \alpha}\frac{\partial T}{\partial t}-\tau_T\frac{\partial^3 T}{\partial t \partial x^2}=0,
\label{temperature_wave_equation}
\end{equation}
with $C$ the volumetric specific heat and $\alpha$=$\kappa_T$/$C$ the thermal diffusivity. Whereas Eq. \ref{temperature_wave_equation} is parabolic, its solution may still bear, under a practical stand-point, wave-like characteristics \cite{Tang1999,ordonez2009thermal} provided $Z$=$\tau_T/\tau_q$$<$1/2 \cite{gandolfi2019accessing}. 
This may be qualitatively understood recognising that the first two terms in Eq. \ref{temperature_wave_equation} constitute the homogeneous wave equation, whereas the last two account for damping effects.

\indent The wave-like propagation characteristics are better appreciated in reciprocal space. Seeking for solutions in the form $T\left(x,t\right)=T(\beta,\omega)\exp\left(i(\beta x+\omega t)\right)$ yields the dispersion relation for the homogeneous case:
\begin{equation}
\beta^2\left(1+i\omega \tau_T\right)=\left( {\tau_q\over \alpha}\right)\omega^2\left(1-\frac{i}{\omega \tau_q} \right),
\label{complex_dispersion}
\end{equation}
linking the angular frequency $\omega$ and wave-vector $\beta$, both quantities being complex-valued. 

\begin{figure}[t]
\begin{center}
\includegraphics[width=0.5\textwidth]{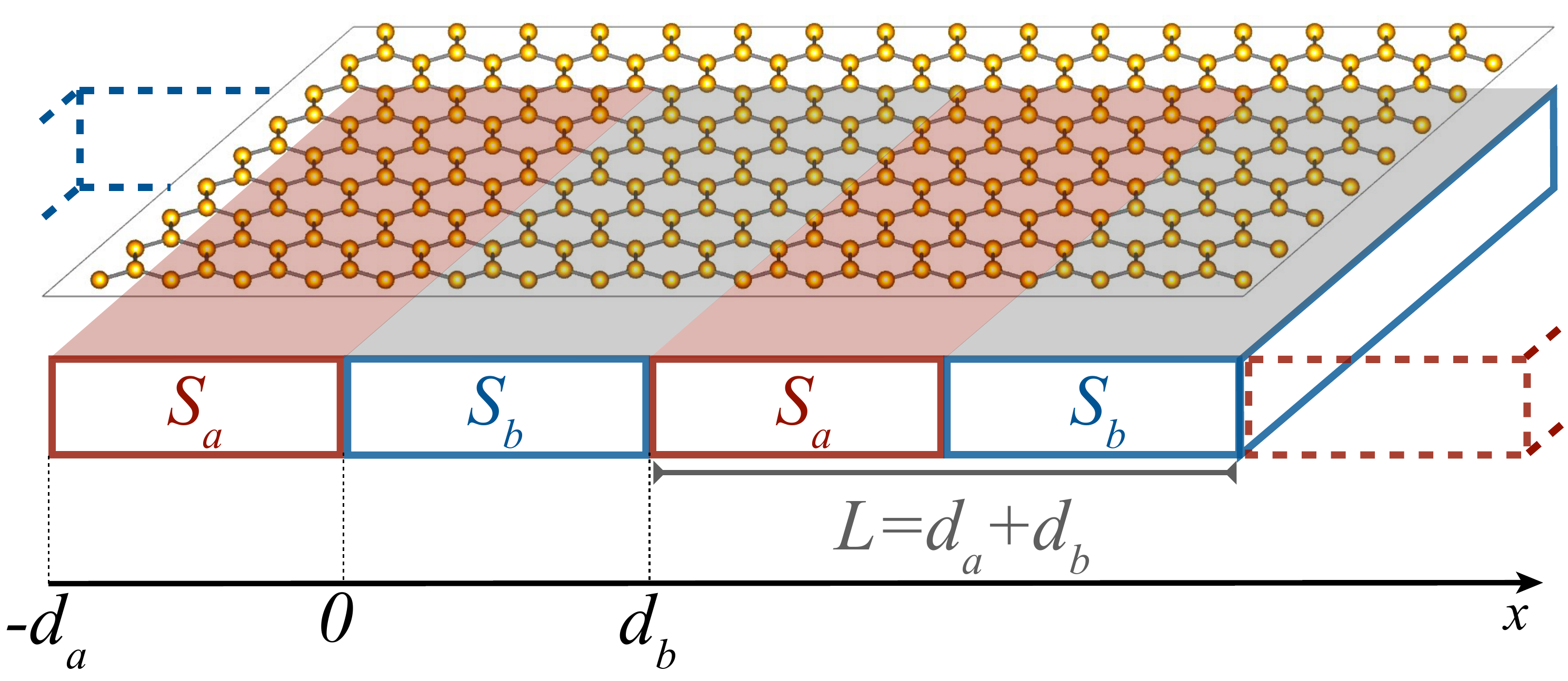} 
\caption{Sketch of the temperonic crystal. Spatially periodic heating of a graphene layer leads to periodic stripes of length $d_a$, $d_b$ and temperature $T_a$, $T_b$, respectively, the u.c. dimension being $L=d_a+d_b$. $S_j$=\{$C$, $\kappa_T$, $\tau_T$, $\tau_q$\}$_{j}$ is the set of thermal properties of stripe $j$, with $j$=\{a,b\}. On a general basis, any periodic structure based on a unit cell composed of two slabs sustaining temperature wave-like oscillations on short time-scales can serve as a TC.}
\label{Sketch_of_geometry}
\end{center}
\end{figure}
Based on the rationale that temperature propagation may preserve coherence properties in a homogenous slab, the concept of a TC is now introduced and its dispersion relation investigated in relation to the feasibility of opening frequencies band gaps in the temperature spectrum. The superlattice u.c., depicted in Fig. \ref{Sketch_of_geometry}, is composed of two slabs $a$ and $b$, each now characterised by the set of thermal parameters $S_{j}$=\{$C$, $\kappa_T$, $\tau_T$, $\tau_q$\}$_{j}$ and length $d_{j}$, the subscript $j$=\{a,b\} denoting the specific slab. The superlattice periodicity is $L=d_{a}+d_{b}$.

Within each slab, the thermal dynamics is described by the set of Eq.s \ref{cons_en_DPLM} upon insertion of the slab's specific thermal parameters. Continuity of temperature and heat flux is enforced at the boundary of each slab.
Furthermore, periodicity is accounted for applying Block-Floquet conditions, $T\left(x=d_b^{-},t\right)=\exp\left(ikL\right)T\left(x=-d_a^{+},t\right)$ and $q\left(x=d_b^{-},t\right)=\exp\left(ikL\right)q\left(x=-d_a^{+},t\right)$, leading to:
$$-{1\over 2}\left[\left(\frac{C_a \beta_b}{C_b \beta_a} \right)+\left(\frac{C_a \beta_b}{C_b \beta_a} \right)^{-1}\right]\sin(\beta_bd_b)\sin(\beta_a d_a)+$$
\begin{equation}
+\cos(\beta_b d_b)\cos(\beta_a d_a)=\cos(kL), \label{Comsol_suca}
\end{equation}
where $\beta_j$ is the wave vector in slab $j$. The $\beta_j$ are functions of $\omega$ via the dispersion relation for the homogeneous case, Eq. \ref{complex_dispersion}. Hence Eq. \ref{Comsol_suca} is the dispersion relation in implicit form, linking the angular frequency $\omega$ and wave vector $k$ of the entire TC. For sake of simplicity, the dispersion relation will be dealt in the frame of the \textit{extended scheme} \cite{ashcroft1976solid}.

In this letter, the interest is in the propagation of a temperature pulse, $\delta T$, along the TC. The temperature pulse is assumed not to substantially affect the TC's temperature, $T_{0}$, that is $\delta T\ll T_{0}$. A temperature pulse may be obtained as a linear superposition of plane waves. With reference to a single plane wave, the present scenario is properly accounted for assuming a $\textit{real-valued}$ $k$ and a $\textit{complex-valued}$ $\omega$=$\omega_{1}$+i$\omega_{2}$,  where $2\pi/|\omega_1|$ and $1/\omega_2$ are the temperature's oscillation period and damping time, respectively \cite{gandolfi2019accessing}. The latter arises as a consequence of the diffusive/damping terms in Eq. \ref{temperature_wave_equation} and sets the time-scale over which wave-like behaviour can be observed.
To quantify the damping of a given mode, the modal quality factor, $Q(k)=|\omega_1(k)|/\omega_2(k)$, is introduced. Numerical solution of Eq. \ref{Comsol_suca} yields $\omega(k)$ and, ultimately, the dispersion curves $\omega_{1}(k)$ and $Q(k)$.

\indent A qualitative understanding of the solution's relevant features may be achieved starting from the dispersion relation of a \textit{homogeneous effective material}, obtained averaging the TC thermal parameters over the u.c. \footnote{
The \textit{homogeneous effective material} is a homogeneous material whom thermal parameters are obtained as an average over the u.c.
In other words, the generic thermal parameter of the homogeneous effective material $X^{eff}$ is obtained as $X^{eff}=(d_{a}X_{a}+d_{b}X_{b})/L$, with $X_{j}$ the corresponding thermal parameter of slab $j$.
}: 
$\omega^{eff}\left(k\right)$=$\omega^{eff}_{1}\left(k\right)$+i$\omega_{2}^{eff}\left(k\right)$, $k$ being real-valued.
Both $\omega^{eff}_{1}$ and the homogeneous effective material quality factor $Q^{eff}$  are continuous versus $k$ \cite{gandolfi2019accessing}, i.e. there are no gaps. Upon adiabatically turning on the periodicity of the thermal parameters, $\omega^{eff}\left(k\right)$ merges into $\omega\left(k\right)$ and frequency gaps in $\omega_{1}$, together with discontinuities in $Q$, open at the Brillouin Zone (BZ) boundaries, that is at wave vectors $k^{BZ}_{n}=\pi n/L$ with $n$ integer.
The band-gap opening is meaningful provided the temperature wave is not overdamped: $Q\left(k^{BZ}_{n}\right)$ (or $Q^{eff}\left(k^{BZ}_{n}\right))>1$ \footnote{
Formally, the criteria $Q^{eff}\left(k^{BZ}_{1}\right)$=$max\{Q^{eff}\left(k\right)\}$ establishes the TC's periodicity enabling \textit{optimal} band-gap opening at the 1$^{st}$ BZ boundary. Nevertheless, the so obtained $L$ may be too short to grant local thermalization, a premise for the DPL model to hold, and/or incompatible with technological constraints.
}, the situation becoming similar to that of standard non-dissipative superlattices \cite{Tamura1988,weisbuch2014quantum}. 
A radically different behaviour occurs for $k\ll k^{BZ}_{1}$, that is for temperature's wavelengths $\lambda\left(=2\pi/k\right)\gg L$. In this case interference effects are negligible and the wave ``feels'' the underlying homogeneous effective material. The TC dispersion is then expected to match that of the homogeneous effective material itself.

We now present a realistic design of a graphene-based TC, where the previous concepts find application for the case of \textit{phononic} temperatures waves \cite{volz2016nanophononics}. Phonons temperature waves were recently reported in graphite at 80 K on the ultrafast time-scale in the seminal work of Huberman et al. \cite{huberman2019observation} and are expected to be sustained also in graphene \cite{cepellotti2020generalization,cepellotti2017transport,cepellotti2016thermal}. The experimental evidences were successfully rationalised by Eq. \ref{complex_dispersion} (with $\omega$ complex and $k$ real), upon identification of $\tau_{T}$ with $\tau_{N}$ and $\tau_{q}$ with $\tau_{U}$, where $\tau_{N}$ and $\tau_{Q}$ are the average phonon scattering times for Normal (N) and Umklapp (U) processes, respectively \cite{gandolfi2019accessing}. Both scattering times are temperature dependent. Onset of the wave-like regime requires $Z$$<$1/2, a condition met at cryogenic temperatures where $\tau_{U}\gg\tau_{N}$. Graphene can thus serve as a suitable building block for a TC operating at temperatures up to 120 K, see Table \ref{tableSummaryCoeffTemperature}, and possibly beyond.

\indent The superlattice is then obtained imposing a static, 1D space-periodic base temperature profile, $T_{0}\left(x\right)$, on a graphene layer, with $T_{0}\left(x\right)$ in the form of a square wave of period $L$, taking the values $T_a$ and $T_b$ over the extents $d_{a}$ and $d_{b}$, respectively \footnote{
A square wave variation of $T_{0}\left(x\right)$ allows a solution in analytical form, thus having the merit of highlighting the relevant physics. In a real case scenario, the spatial temperature modulation would though be smoother. This may possibly lead to a reduction of the band gaps frequency extent occurring between \textit{high order} frequency bands, their frequency position remaining primarily determined by the TC periodicity. 
}.
Both $T_a$ and $T_b$ are chosen so as to ensure that N-scattering events prevail over U ones (hence $Z\ll1$). $T_{0}\left(x\right)$ leads to a periodic modulation of the graphene's sheet thermal parameters through their temperature dependence, ultimately resulting in the TC depicted in Fig. \ref{Sketch_of_geometry}. Practically, $T_{0}\left(x\right)$ can be applied via a space-periodic temperature modulation of a supporting chip on which the graphene layer adheres.
This can be achieved in a variety of ways: for instance by exploiting the photo-thermal effect and shining an optical grating on a homogeneous substrate material kept at cryogenic temperature. Another possibility is by homogeneous illumination of a cryogenic substrate made of stripes with different absorption coefficients or via nanopatterning a periodic resistor array on the supporting substrate.

\indent A specific case is now addressed, in which the u.c. slabs $a$ and $b$ are kept at temperature $T_a=100$ K and $T_b=80$ K, respectively. For sake of simplicity, in what follows a symmetric u.c. is assumed: $d_a$=$d_b$=$L/2$. Furthermore, graphite's thermal properties are taken (see Table \ref{tableSummaryCoeffTemperature}) as a good compromise for the wide range of graphene's thermal properties reported in the literature, the latter strongly depending on a plethora of parameters ranging from degree of crystallinity, production method, grain size to the kind of supporting material \cite{nika2009lattice,kolesnikov2012low,seol2010,pop2012}.
The values for $\tau_T$ and $\tau_q$ have been adopted \cite{gandolfi2019accessing} upon fitting the experimental dispersion relation  \cite{huberman2019observation} with Eq. \ref{complex_dispersion}, i.e. within the frame of the DPL. These values are consistent with the scattering times obtained in the frame of the Boltzmann Transport Equation \cite{cepellotti2015phonon}.
The focus will be limited to opening of the first band gap, $n$=1, the discussion being similar for higher gaps.

\begin{figure}[t]
\begin{center}
\includegraphics[width=0.4\textwidth]{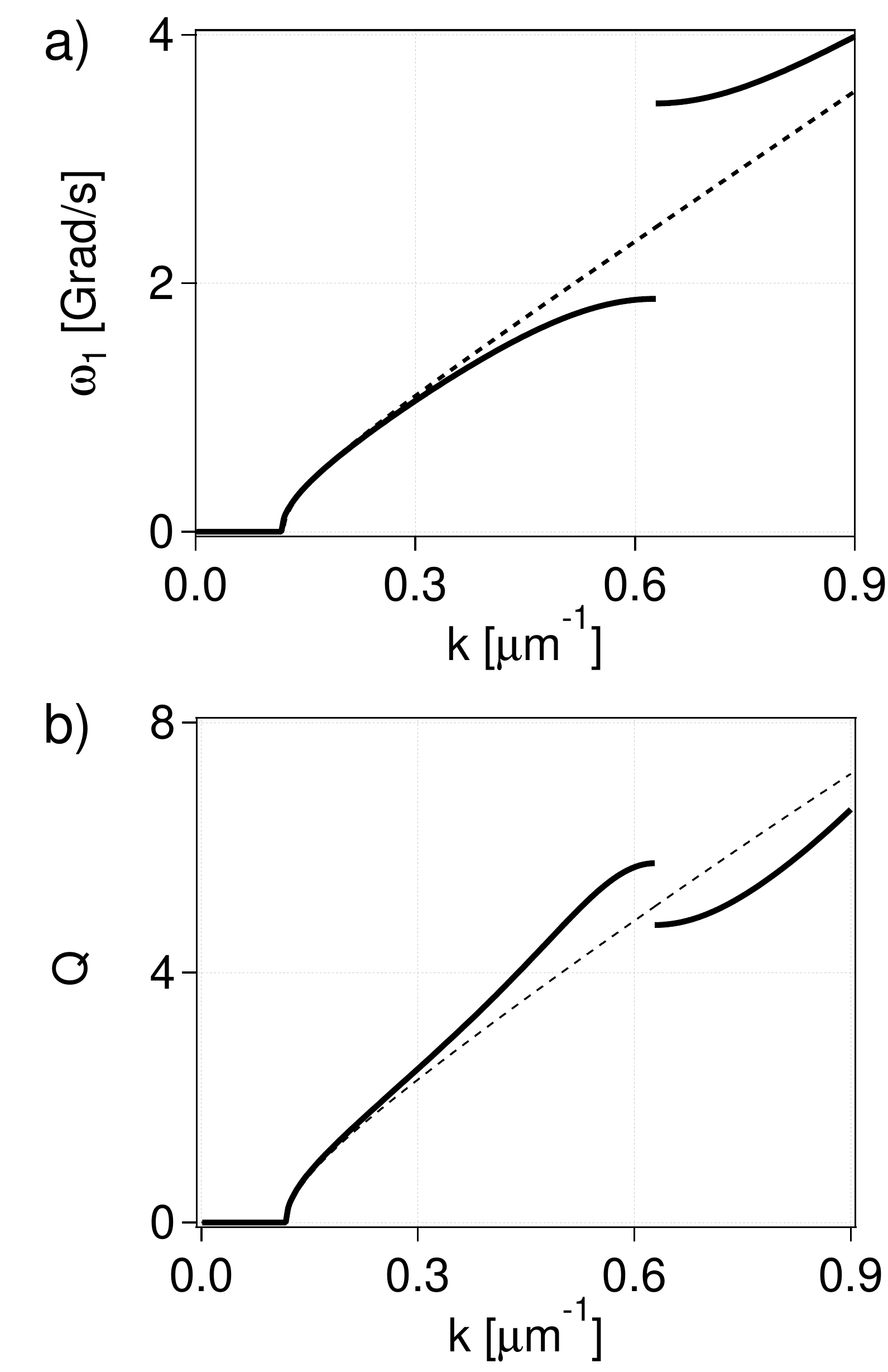} 
\caption{Temperature band structure: dispersion relation $\omega_1$ (panel a) and quality factor Q (panel b). Black full line: graphene TC with $T_a=100$ K and $T_b=80$ K. Black dashed line: $\omega^{eff}_{1}$ (panel a) and $Q^{eff}$ (panel b) for the corresponding homogeneous effective material. The unit cell size is $5\ \um{\mu m}$.}
\label{Dispersion_and_Q}
\end{center}
\end{figure}

\begin{table}[h]
\centering
\begin{tabular}{|c|c|c|c|c|} 
 \hline  
Thermal&	$T=80$ K & $T=100$ K&$T=110$ K &$T=120$ K \\ 
parameter&&&&\\ \hline
$\kappa_T\  [\um{W\over m\ K}]$ \cite{fugallo2014thermal}& $4.3\times 10^3$ & $5.0\times 10^3$& $5.0\times 10^3$&$5.0\times 10^3$\\ \hline
$C\  [\um{J\over m^3 K}]$ \cite{desorbo1953specific} & $2.3\times 10^5$ & $3.3\times 10^5$ 	&$3.7\times 10^5$ &$4.4\times 10^5$ \\ \hline
$\tau_T\  [\um{ps}]$ \cite{gandolfi2019accessing,cepellotti2015phonon}& $3.0$ & $2.5$ &$2.2$&$2.1$ \\ \hline
$\tau_q\  [\um{ps}]$  \cite{gandolfi2019accessing,cepellotti2015phonon}& $1800$& $300$&$140$&$90$  \\ \hline
\end{tabular}
\caption{Thermal parameters for graphene as a function of temperature. The temperatures have been chosen so as to fulfil the requirements for wave-like propagation: $Z<1/2$.}
\label{tableSummaryCoeffTemperature}
\end{table}

The black dashed line in Fig. \ref{Dispersion_and_Q} represents $\omega^{eff}_{1}$ (panel a) and $Q^{eff}$ (panel b) vs $k$ for the \textit{homogeneous effective material}.
The u.c. size, $L$, is chosen to be 5 $\mu$m, similarly to the transient grating periodicity exploited in Ref.\cite{huberman2019observation} to observe temperature waves in graphite at 80 K.
The black full lines in Fig. \ref{Dispersion_and_Q} represent $\omega_1$ (panel a) and $Q$ (panel b) vs $k$ for the TC. In particular, a gap opens in the $\omega_{1}$ spectrum, implying that temperature waves with $\omega_{1}$ in the 1.8 to 3.5 Grad/s range are not supported by the TC.
The gap extent, $\Delta_g$, amounts to 1.7 Grad/s and is centered at $\sim\omega^{eff}_{1}\left(k^{BZ}_{1}\right)\sim$ 2.5 Grad/s, the latter being the value of $\omega^{eff}_1$ at the 1$^{st}$ BZ boundary. The periodicity is very effective in opening the band gap, the relevant figure of merit (FOM) being $\Delta_{g}/|\omega^{eff}_{1}\left(k^{BZ}_{1}\right)|\sim\ 64\%$. The TC hence works as an effective frequency filter for temperature waves operating on a time-scale $\sim Q^{eff}\left(k^{BZ}_{1}\right)/|\omega_{1}\left(k^{BZ}_{1}\right)|\sim$ 2 ns.

For temperature's wavelengths $\lambda\gg L$ ($k\ll 2k^{BZ}_{1}$) the TC dispersion relation matches that of the homogeneous effective material, as appreciated inspecting Fig. \ref{Dispersion_and_Q}  for $k\lesssim 0.3\ \um{\mu m}^{-1}$.
Within this regime one can thus engineer an artificial material sustaining temperature waves on a time scale $\sim Q^{eff}\left(k\right)/|\omega_{1}\left(k\right)|$, which effective thermal properties are obtained as an average over the u.c.

We now investigate the gap tunability in terms of spectral position and extent, upon varying the slabs thermal properties via their temperature dependence. To this end, and following the same rationale as for the previous case, we report results for two additional TCs obtained keeping $T_a$=100 K while augmenting $T_{b}$ to 110 K and 120 K, respectively.
The u.c. size is chosen to be $L=2.1\ \um{\mu m}$ and $L=2\ \um{\mu m}$ respectively. On a general basis, augmenting the working temperature diminishes the phonon scattering times, hence allowing for a shorter $L$, provided the requirement $Z$$<$1/2 still holds. These periodicities are readily implementable with current nanopatterning techniques.

\begin{figure}[t]
\begin{center}
\includegraphics[width=0.39\textwidth]{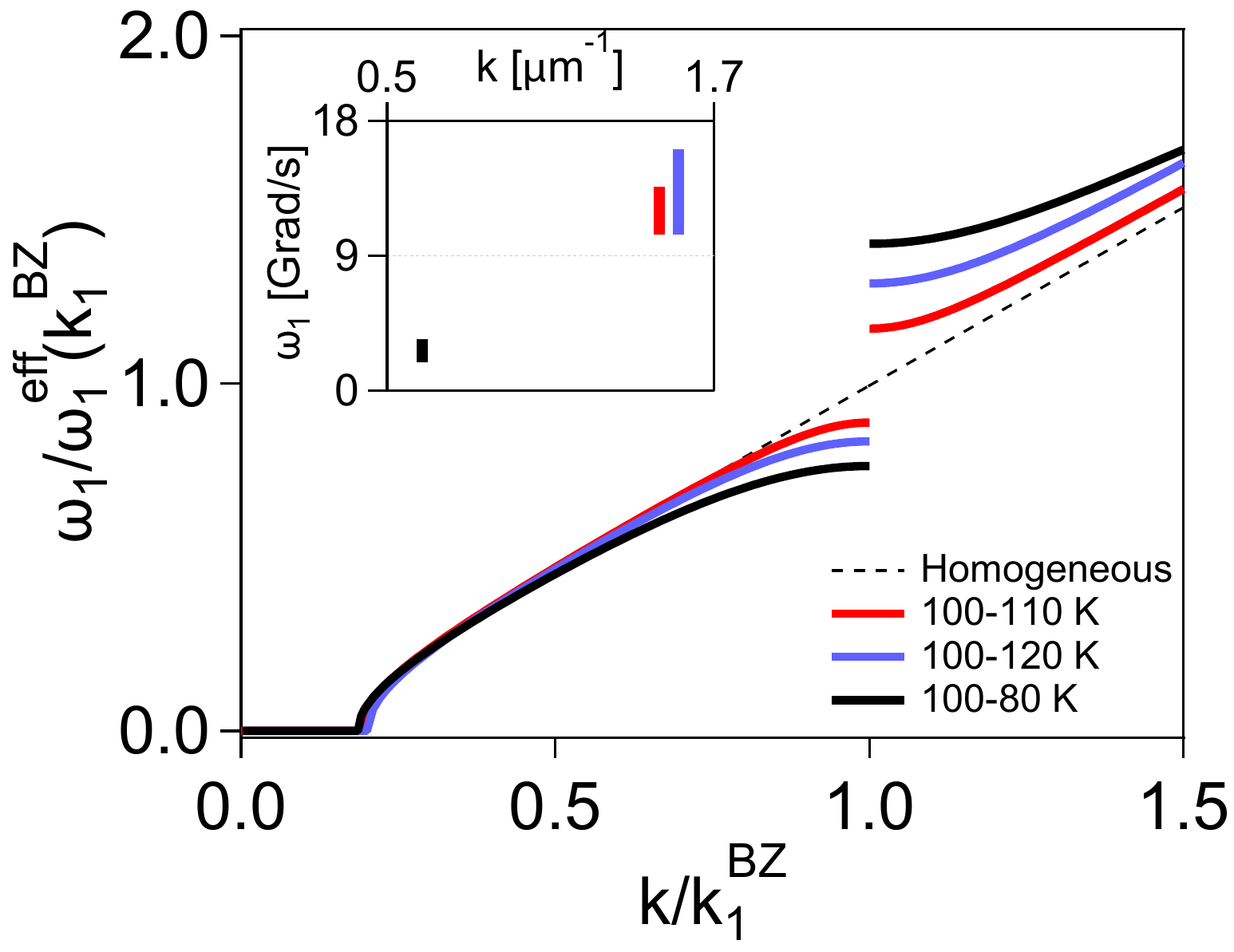} 
\caption{\textit{Normalized} dispersion relations for a graphene TC with $T_a=100$ K kept fixed and $T_b=110$ K (red full line), 120 K (blue full line), 80 K (black full line). Black dashed line: \textit{normalized} dispersion relation for the corresponding homogeneous effective materials (same curve for all three cases).
Inset: 1$^{st}$ gap location and extent (vertical bars) for each TC in actual, i.e. non-normalized, units.}
\label{Normalized_dispersion}
\end{center}
\end{figure}

Fig. \ref{Normalized_dispersion} shows the normalized dispersion relations for all three TCs. For sake of comparison, $k$ and $\omega_{1}$ have been normalized against $k^{BZ}_{1}$ and $\omega^{eff}_{1}\left(k^{BZ}_{1}\right)$, the latter two values being specific to each TC. The $\textit{normalized}$ dispersion relations of the three homogeneous effective materials are thus the same (black dashed line). The normalized band gaps correspond to the FOMs. As the temperature difference, $|T_b-T_a|$, augments for a fixed value of $T_a$, so does the FOM, which is $27\%$, $46\%$ and $64\%$ for values of $T_b$=110 (red line), 120 (blu line) and 80 K (black line), respectively. Mind though that the FOM does not depend on the temperature difference only, but also on the actual slabs temperatures. For instance, $|T_b-T_a|$= 20 K for the TCs with $T_b$=80 K (full black line) and 120 K (full blue line) but the FOMs are different.
The actual (i.e. non-normalized) band gaps are shown as vertical bars in inset of Fig. \ref{Normalized_dispersion}. What emerges is that the band gap can be tuned, both in spectral position and extent, over a wide range of angular frequencies, all within reach of state-of-the art optothermal spectroscopies \cite{gandolfi2017emergent,huberman2019observation,beardo2020}.\\
The Temperonic Crystal, a superlattice for temperature wave-like motion occurring in the hydrodynamic regime, was introduced. The band structure for temperature waves may be engineered tuning the TC thermal and geometrical properties. The concept was illustrated for the case of a graphene-based TC, where temperature frequency gaps in the 1-10 Grad/sec range and Q-factor $\sim$ 5 were obtained for TC periodicities $\sim$ $5\ \um{\mu m}$ and working temperatures above liquid nitrogen's. Tuning of the temperature band structure was demonstrated under realistic conditions.

The TC concept can be readily expanded to encompass other materials sustaining temperature waves in the hydrodynamic regime such as graphite and single-layer transition metal dichalcogenides \cite{Torres2019}, for the case of phononic temperature, and quantum materials \cite{gandolfi2019accessing,gandolfi2017emergent}, for the case of electronic and spin temperatures. For the latter two cases (quantum materials) the scattering times can be faster than the ones presently addressed, potentially allowing for a shorter TC's periodicity, enhanced Q-factors and requiring observation on faster time scales.
TC based on two dimensional materials are particularly amenable for integration on nanostructured supports \cite{Gu2015Phonon}. In this context twisted-graphene superlattices \cite{cao2018unconventional,cao2018correlated} merge the properties of graphene and correlated materials offering the possibility of controlling the ratio between the local interaction strength and the effective bandwidth.
This could provide enhanced thermal parameters tunability upon acting on the magic-angle of twisted bilayer graphene.

The TC may be achieved also with alternate layout designs. For instance, one could envision creating the TC alternating two materials with a good contrast in their mutual thermal properties instead of imposing the periodicity via a thermal modulation of the same material.

In persepctive, manipulation of temperature pulses on short time and length scales, by exploiting interference effects as the one here shown, opens up to thermal nanodevice concepts based on coherence effects such as waveguides, cavities and frequency filters for the temperature field.

\section*{Acknowledgements}
M.G. acknowledges financial support from the CNR Joint Laboratories program 2019-2021. C.G. acknowledges support from Universit\`a Cattolica del Sacro Cuore through D.2.2 and D.3.1 grants, from MIUR through PRIN 2015 (Prot 2015C5SEJJ001) and PRIN 2017 (Prot. 20172H2SC4$\_$005) programs. F.B. acknowledges financial support from Universit\'e de Lyon in the frame of the IDEXLYON Project (ANR-16-IDEX-0005) and from Universit\'e Claude Bernard Lyon 1 through the BQR Accueil EC 2019 grant. 

\bibliography{Gandolfi_references}

\end{document}